\documentclass[AMA]{WileyNJD-v1}
\articletype{Original Research Article}%

\usepackage{multirow}
\usepackage[superscript,biblabel]{cite}

\newenvironment{lyxlist}[1]
{\begin{list}{}
{\settowidth{\labelwidth}{#1}
 \setlength{\leftmargin}{\labelwidth}
 \addtolength{\leftmargin}{\labelsep}
 }}
{\end{list}}

\begin{document}


\title{Quantifying and Estimating the Predictive Accuracy for Censored Time-to-Event
Data with Competing Risks}

\author[1,2]{Cai Wu}

\author[2]{Liang Li*}

\authormark{ C. WU AND L. LI}

\address[1]{\orgdiv{Department of Biostatistics}, \orgname{The University of Texas Health Science Center at Houston}, \orgaddress{\state{TX}, \country{USA}}}

\address[2]{\orgdiv{Department of Biostatistics}, \orgname{The University of Texas MD Anderson Cancer Center}, \orgaddress{\state{TX}, \country{USA}}}

\corres{*Liang Li, Department of Biostatistics, The University of Texas MD Anderson Cancer Center, Houston, TX 77054. \email{lli15@mdanderson.org}}

\abstract{This paper focuses on quantifying and estimating the predictive accuracy
of prognostic models for time-to-event outcomes with competing events.
We consider the time-dependent discrimination and calibration metrics,
including the receiver operating characteristics curve and the
Brier score, in the context of competing risks. To address censoring,
we propose a unified nonparametric estimation framework for both discrimination
and calibration measures, by weighting the censored subjects with the conditional probability of the event
of interest given the observed data.
 We demonstrate through simulations
that the proposed estimator is unbiased, efficient and robust against
model misspecification in comparison to other methods published in
the literature. In addition, the proposed method can be extended to
time-dependent predictive accuracy metrics constructed from a general
class of loss functions. We apply the methodology to a data set from the African American Study of Kidney Disease and Hypertension to
evaluate the predictive accuracy of a prognostic risk score in predicting
end-stage renal disease (ESRD), accounting for the competing risk of pre-ESRD death.}


\maketitle

\section{Introduction}

\renewcommand\[{\begin{equation}}

\renewcommand\]{\end{equation}} 

In modern evidence-based medicine, decisions on a diagnosis or personalized
treatment plan are often guided by risk scores generated from prognostic
models \cite{baskin2007recipient,hernandez2009novel,lorent2016mortality}.
Such prognostic risk scores can be either a single risk factor, such as a biomarker, or a risk probability calculated from multiple
risk factors. For a risk score to be utilized in clinical practice,
its predictive accuracy is often assessed through two types of metrics:
(1) the discrimination metric, which measures how well the risk score
can distinguish subjects with and without the disease condition,
and (2) the calibration metric, which measures how well the predicted
risk matches the observed risk in the target population. Motivated
by the prediction of end-stage renal disease (ESRD) among a cohort of patients with chronic kidney disease, the goal
of this paper is to propose a framework to estimate the predictive
accuracy of a risk score from a prognostic model, accounting
for right censoring and competing events.

For a continuous time-to-event outcome, the presence and absence
of a disease condition at any time point $\tau$ can be viewed as a binary outcome. To study the relationship between a continuous risk
score and this binary outcome at any prespecified time point $\tau$, the
time-dependent receiver operating characteristics (ROC) curve is widely used for assessing discrimination, i.e., the separation of subjects with and without a given disease at time $\tau$ by the risk score \cite{heagerty2000time}. For example, the risk score is the $\tau$-year (e.g., $\tau=5$) survival probability
calculated based on the characteristics of a cancer patient at initial
diagnosis, and the disease presence or absence is defined by whether
the patient died of cancer within $\tau$ years after the initial
diagnosis. For such a risk score, the area under the ROC curve (AUC)
presents the probability that a subject with the disease at time $\tau$ has a higher predicted risk score than a subject without the disease. A challenge of estimating such time-dependent ROC curve is that the disease
status at $\tau$ is unknown among subjects who are censored prior to $\tau$. A number of methods
have been developed
to address this issue, including the nearest neighboring estimator (NNE)
\cite{heagerty2000time} and inverse probability censoring weighting
(IPCW) \cite{blanche2013review,chiang2010non,uno2007evaluating}.
In addition to the metrics for discrimination, metrics for calibration \cite{graf1999assessment} quantify the absolute
deviance of the risk score from the observed outcome, known as the prediction error. Time-dependent prediction error
metrics for survival outcomes have been proposed \cite{graf1999assessment,gerds2006consistent,korn1990measures,schemper2000predictive}.
The prediction error can be constructed through a class of
loss functions that link the risk score and the binary disease outcome
at $\tau$ \cite{graf1999assessment}. Among those, the quadratic loss, known as the Brier score
\cite{brier1950verification}, is a popular choice \cite{blanche2015quantifying,cortese2013comparing,parast2012landmark}. Censoring remains a challenge when estimating the Brier score, and an IPCW method
was proposed to deal with it \cite{graf1999assessment,gerds2006consistent}.

Competing risks are common in clinical research that involves time-to-event data. For example, in a cardiovascular study, one may be interested in the time to the first myocardial infarction after cardiovascular surgery, but patients may die before experiencing the event of interest. Limited statistical methodology is available to estimate the predictive accuracy metrics in the context of competing risks. To estimate the time-dependent
ROC,  Saha \& Heagerty \cite{saha2010time} extended the NNE method \cite{heagerty2000time}
to the competing risk context. Zheng et al. \cite{zheng2012evaluating} further
extended the method of  Saha \& Heagerty \cite{saha2010time} to covariate-adjusted time-dependent
ROC. Blanche et al. \cite{blanche2013estimating} studied the use of IPCW in estimating the time-dependent ROC with competing risk data.
For the estimation of the Brier score with competing risk data, the available published methods are based on the IPCW \cite{blanche2015quantifying,liu2016robust,schoop2011quantifying}, with the censoring distribution estimated either by the Kaplan-Meier (KM) method without conditioning on the risk score \cite{graf1999assessment} or by the Cox proportional hazards model conditional on the risk score \cite{gerds2006consistent}.

This paper focuses on the time-dependent discrimination and calibration
estimation in the context of competing risk outcomes. We propose a novel
nonparametric kernel-weighted estimation framework for both time-dependent discrimination and calibration measures. The proposed method first estimates the conditional probability of experiencing an event of interest at $\tau$
given the observed data of the subjects. This is done through nonparametric
kernel regression for the cumulative incidence function. Then the
time-dependent predictive accuracy metrics, such as sensitivity, specificity,
and Brier score, are estimated by weighting each subject with their
own conditional probabilities.

The proposed method has some attractive properties. First, it is fully
nonparametric, without any distributional or modeling assumptions.
This is desirable for estimating predictive accuracy metrics since
it reduces the bias from the estimation procedure itself. Second, the
proposed method, unlike other nonparametric methods such as NNE
\cite{heagerty2000time}, is insensitive to the bandwidth choice. This
is shown in this paper with both numerical and methodological
justifications. Third, the method automatically accommodates correlation between
the censoring time and the risk score. Furthermore, the proposed method can be invariant
to monotone transformation of the risk score when the tuning parameter
is specified by the span, the proportion of subjects included in the kernel estimation.
Also, the estimated sensitivity, specificity, and ROC curve are monotone in
the cut-off point $c$. 
Our simulation shows that the proposed method has competitive performance
in terms of bias and the mean squared error (MSE) when compared with other published
methods. Section 2 presents the notations and definitions for the time-dependent
ROC and time-dependent prediction error. Section 3 describes the proposed
estimators for the predictive accuracy metrics. Then the finite sample
performance is evaluated by simulations in Section 4. In Section
5, we illustrate the method with data from the African American Study of Kidney
Disease and Hypertension (AASK) in evaluating the prediction
of ESRD. Section 6 concludes the paper by discussing the findings and
providing some perspective.

\section{Predictive Accuracy for Time-to-Event Data with Competing Risks}

\subsection{Notation}

Let $T$ denote the event time, $C$ the censoring time, $\delta$
the event type, and $\Delta=1(T\le C)$ the censoring indicator, where
$1(\cdot)$ is the indicator function. We observe independent and
identically distributed (i.i.d.) samples of $\{(\tilde{T}_{i},U_{i},\tilde{\delta}_{i}),i=1,2,\ldots n\}$ in a validation data set,
where $\tilde{T}_{i}=\textrm{min}(T_{i},C_{i})$ is the observed
time to the event or censoring, whichever comes first. The observed status
$\tilde{\delta}_{i}=\Delta_{i}\delta_{i}$, which equals zero for censored
subjects and equals one of the $K$ possible causes, $\delta_{i}\in\{1,2,\ldots K\}$,
for uncensored subjects. Without loss of generality, we present our
methodology with $K=2$ to match the data application in Section
5. The methodology still applies with other choices of $K\ (K>2)$. For clarity,
suppose that we are interested in assessing the predictive accuracy
of event type $\delta=1$. Let $U_{i}$ denote the risk
score for subject $i$, with higher values of $U_{i}$ indicating higher
risk of the event. For example, $U_{i}$
can be the predicted cumulative incidence probability from a competing
risk regression model that we want to evaluate, i.e., $U_{i}=\pi_{1}(\tau\vert\boldsymbol{Z}_{i})=P(T_{i}\le\tau,\delta_{i}=1\vert\boldsymbol{Z}_{i})$, where $\boldsymbol{Z}$ denotes the predictor and $\tau$ is the predictive horizon. The predictive model is often developed from a training data set that is different from the validation data set. 
This paper focuses on estimating the predictive accuracy metrics in a validation data set. We do not study how the model for the risk score $U$ is estimated or whether the model is correctly estimated. We assume that this model has already been developed, needs to be evaluated, and the risk score $U$ has the interpretation of being the subject-specific predicted cumulative incidence probability at horizon $\tau$.

\subsection{Definitions of the time-dependent ROC curve and AUC\label{subsec:tdROC_comprsk}}

In the presence of competing events, the definition of cases is straightforward.
The cases at time $\tau$ for event type $k$ are defined as subjects
who undergo event $\delta=k$ before time $\tau$, i.e., $Case_{k}=\{i:T_{i}\le\tau,\delta_{i}=k\}$.
At a given threshold $c$, the cause-specific sensitivity at time
$\tau$ is defined as
\begin{equation}
Se(c,\tau)=P(U>c\vert T\le\tau,\delta=k).\label{eq:def_sen}
\end{equation}
This is the definition of $cumulative/dynamic$ sensitivity
\cite{heagerty2000time}. When $U$ is higher than the threshold value
$c$, the patient is predicted to experience event $k$ within the time
window $(0,\tau]$. 

We consider two definitions of controls that lead
to two different definitions of time-dependent specificity. Saha \& Heagerty \cite{saha2010time} originally
defined the control group at time $\tau$ as the event-free subjects,
i.e., $\{i:T_{i}>\tau\}$. According to this definition, subjects who
experienced competing events other than $k$ are neither cases nor
controls. Therefore, Zheng et al. \cite{zheng2012evaluating} introduced an alternative
definition of the control group $\{i:T_{i}>t\}\cup\{i:T_{i}\le t,\delta_{i}\ne k\}$,
which includes both event-free subjects and subjects who experience other competing
events. We study the estimation under both definitions:
\begin{lyxlist}{00.00.0000}
\item [{\textbf{Definition}}] \textbf{A. }Case $k$: $T\le\tau,\delta=k$;
Control$_{A}$: $(T>\tau)\cup(T\le\tau\cap\delta\ne k).$
\end{lyxlist}
\begin{lyxlist}{00.00.0000}
\item [{\textbf{Definition}}] \textbf{B.} Case $k$: $T\le\tau,\delta=k$;
Control$_{B}$: $T>t$ . 
\end{lyxlist}
The specificity at time $\tau$ with respect to the two types of definitions
is 
\begin{align}
Sp_{A}(c,\tau) & =P(U\le c\vert\{T>\tau\}\cup\{T\le\tau,\delta\ne k\})\nonumber \\
Sp_{B}(c,\tau) & =P(U\le c\vert T>\tau).\label{eq:def_sp}
\end{align}
Two different time-dependent ROC curves can be obtained by plotting
$Se(c,\tau)$ versus either $1-Sp_{A}(c,\tau)$ or $1-Sp_{B}(c,\tau)$,
i.e., $ROC_{A}(x,\tau)=Se(Sp_{A}^{-1}(1-x,\tau),\tau)$ and $ROC_{B}(x,\tau)=Se(Sp_{B}^{-1}(1-x,\tau),\tau)$
for $x\in[0,1]$. The corresponding $AUC$s can be defined as $AUC(\tau)=\int_{0}^{1}ROC(x,\tau)dx$ or
as the proportion of concordance pairs among the population \cite{blanche2013estimating}:
\begin{align}
AUC_{A}(\tau) & =P(U_{i}>U_{j}\vert T_{i}\le\tau,\delta_{i}=k,\{T_{j}>\tau\}\cup\{T_{j}\le\tau,\delta_{j}\ne k\})\nonumber \\
AUC_{B}(\tau) & =P(U_{i}>U_{j}\vert T_{i}\le\tau,\delta_{i}=k,T_{j}>\tau),\label{eq:AUC def}
\end{align}
where $i$ and $j$ indicate two independent subjects under comparison.
The subjects who experienced the competing events before $\tau$ contribute
to $AUC_{A}(\tau)$ but not $AUC_{B}(\tau)$. The justification for
both definitions is related to the clinical interpretation \cite{zheng2012evaluating}. 

\subsection{Definitions of the time-dependent prediction error }
The time-dependent prediction error in the competing risk framework
is defined as the distance between the event-specific status $1(T\le\tau,\delta=k)$
and the subject-specific predicted cumulative incidence function  at horizon $\tau$, $\pi_{k}(\tau\vert \boldsymbol{Z})=P(T\le\tau,\delta=k\vert\boldsymbol{Z})$.
Suppose we are interested in evaluating the prediction for event type
1, three types of prediction error measurements can be defined as
follows \cite{van2011dynamic}:
\[
AbsErr(\tau)=E\Big|1\{T\le\tau,\delta=1\}-\pi_{1}(\tau\vert\mathbf{Z})\Big|
\]
\[
Brier(\tau)=E\Big[1\{T\le\tau,\delta=1\}-\pi_{1}(\tau\vert\mathbf{Z})\Big]^{2}
\]
\[
KL(\tau)=-E\Big[1\{T\le\tau,\delta=1\}\cdot\textrm{ln}\pi_{1}(\tau\vert\mathbf{Z})+1\{(T>\tau)\cup(T\le\tau,\delta\ne1)\}\cdot\textrm{ln(}1-\pi_{1}(\tau\vert\mathbf{Z}))\Big].
\]
Among the three measures, $AbsErr(\tau)$ is not \textquotedblleft proper\textquotedblright{}
in the sense that it is not minimized by the predicted cumulative incidence function (CIF) from the true model \cite{graf1999assessment}.
$Brier(\tau)$ is not only \textquotedblleft proper\textquotedblright,
but has the attractive property that it can be decomposed into a term related to the
bias of the predictive survival probability and a term related to the variance of disease status \cite{schoop2011quantifying}.
The Kullback-Leibler score, $KL(\tau)$, has a close connection to the
likelihood ratio test and the Akaike information criteria (AIC), but its
disadvantage is that $KL(\tau)$ goes to infinity when $\pi_{1}(\tau\vert\mathbf{Z})=0$
and $\{T\le\tau,\delta=1\}$, or when $\pi_{1}(\tau\vert\mathbf{Z})=1$
and $\{T>\tau$ or $T\le\tau,\delta\ne1\}$ \cite{van2011dynamic}. The Brier score is more widely used than the other two, and we will
focus on the Brier score for the rest of this paper, even though our methodology also applies to the other two metrics.

\section{The Proposed Nonparametric Weighting Estimators}

Without censoring, sensitivity and specificity can be estimated
empirically as the fraction of true positives and true negatives.
However, when subjects are censored before $\tau$, the true disease
status at $\tau$ is unknown. The empirical fractions can no longer
be used and proper adjustment for censoring is needed. In the context
of right-censored data without competing events, Li et al. \cite{li2016simple} proposed to weigh each
subject by their respective conditional probability of having the
disease at $\tau$ given all the observed data for that subject. The
conditional probability equals 0 if a subject survives beyond $\tau$
without the disease or 1 if the subject acquires the disease prior to $\tau$.
If a subject is censored prior to $\tau$, the conditional probability
is estimated through a nonparametric kernel regression. In this paper,
we extend that approach to the context of competing risk data. The weight
is defined as the conditional probability of being a case prior to
time $\tau$ given the observed time to the event, event status and prognostic
risk score: 

\begin{align}
W_{1i} & =P(T_{i}\le\tau,\delta_{i}=1\vert\tilde{T}_{i},\tilde{\delta}_{i},U_{i})\nonumber \\
 & =\Big\{1(\tilde{\delta}_{i}=0)\cdot\frac{F_{1}(\tau\vert U_{i})-F_{1}(\tilde{T}_{i}\vert U_{i})}{S(\tilde{T}_{i}\vert U_{i})}+1(\tilde{\delta}_{i}=1)\Big\}\cdot1(\tilde{T}_{i}\le\tau),\label{eq:wt_cmprsk}
\end{align}
where $F_{1}(t\vert U_{i})=P(T_{i}\le t,\delta_{i}=1\vert U_{i})$ is the conditional cumulative incidence
function for event 1, and $S(t\vert U_{i})=P(T_{i}>t \vert U_{i})$ is the conditional overall survival probability.
According to equation (\ref{eq:wt_cmprsk}), we have $W_{1i}=1$ for
subjects with observed event 1 before $\tau$: $\{i:\tilde{T}_{i}\le\tau,\tilde{\delta}_{i}=1\}$;
$W_{1i}=0$ for subjects without any events before $\tau$ or with
competing events before $\tau$: $\Big\{ i:\{\tilde{T}_{i}>\tau\}\cup\{\tilde{T}_{i}\le\tau,\tilde{\delta}_{i}\notin\{0,1\}\}\Big\}$
; and $W_{1i}=\dfrac{F_{1}(\tau\vert U_{i})-F_{1}(\tilde{T}_{i}\vert U_{i})}{S(\tilde{T}_{i}\vert U_{i})}$
for subjects censored before $\tau$: $\{i:\tilde{T}_{i}\le\tau,\tilde{\delta}_{i}=0\}$.
This weighting approach uses the observed status for uncensored subjects and only
imputes the unknown status for censored subjects with a probability.
A heuristic justification is that the case group includes not
only those who are known to have experienced event 1 but also fractions
of those whose status is unknown due to censoring. Similar justification applies to the controls. This differs from
the IPCW method \cite{blanche2013estimating,schoop2011quantifying}, which uses only uncensored subjects and reweights them to account for 
censoring. The IPCW weight is defined as $W_{i}^{IPCW}(\tau)=\dfrac{1(T_{i}\le\tau,\tilde{\delta}_{i}\ne0)}{G_{n}(\tilde{T_{i}}\vert \cdot)}+\dfrac{1(T_{i}>\tau)}{G_{n}(\tau\vert \cdot)}$. It is 
the inverse of the probability of being censored, where $G(t\vert \cdot)$
is the censoring distribution that can be estimated by the Kaplan-Meier
estimator or conditionally given covariates.

Estimation of the proposed weight (\ref{eq:wt_cmprsk}) includes estimation
of two quantities: the conditional CIF $F_{1}(\cdot\vert U_{i})$
and the conditional overall survival probability $S(\cdot\vert U_{i})$. We propose
to use a nonparametric kernel-weighted Kaplan-Meier estimator \cite{li2016simple}:

\begin{equation}
\widehat{S}_{T}(t\vert U_{i})=\prod_{\zeta\in\Omega,\zeta\le t}\Big\{1-\frac{\sum_{j}K_{h}(U_{j},U_{i})\cdot1(\tilde{T}_{j}=\zeta,\tilde{\delta}_{j}\ne0)}{\sum_{j}K_{h}(U_{j},U_{i})\cdot1(\tilde{T}_{j}\ge\zeta)}\Big\},\label{eq:kernel_surv}
\end{equation}

and the kernel-weighted CIF \cite{kalbfleisch2011statistical}:

\begin{equation}
\widehat{F}_{1}(t\vert U_{i})=\sum_{\zeta\in\Omega,\zeta\le t}\frac{\sum_{j}K_{h}(U_{j},U_{i})1(\tilde{T_{j}}=\zeta,\tilde{\delta_{j}}=1)}{\sum_{j}K_{h}(U_{j},U_{i})1(\tilde{T}_{j}\ge\zeta)}\cdot\widehat{S}_{T}(\zeta-\vert U_{i}).\label{eq:kernel_cif}
\end{equation}
$\Omega$ is the set of distinct $\tilde{T}_{i}$'s for $\tilde{\delta_{j}}\ne0$
; and $K_{h}(x,x_{0})=\frac{1}{h}K(\frac{x-x_{0}}{h})$ is the kernel
weight with kernel function $K(\cdot)$ and bandwidth $h$. Alternatively,
we can specify a $span$ instead of a fixed bandwidth. A span is the
proportion of subjects around the neighborhood involved in the kernel
estimation with a uniform kernel function. In implementation, the CIF in (\ref{eq:kernel_cif}) can be estimated as a Kaplan-Meier type product-limit estimator, with the hazard function being replaced by the sub-distribution hazard. The at-risk set in the sub-distribution hazard is obtained by reweighting the individuals who had competing events.  This process can be achieved by reformatting the competing risk data into a counting
process with \texttt{crprep()} function from the \texttt{mstate} package, and using \texttt{survfit()} in the \texttt{survival}
package by specifying a time-dependent $weight$ in \texttt{R} \cite{geskus2011cause}. 

\subsection{The proposed weighting estimators for the time-dependent ROC curve and
AUC }

The estimated weight $\widehat{W}_{1i}$ can be obtained by replacing the CIF and survival functions in (\ref{eq:wt_cmprsk}) with their estimators given by (\ref{eq:kernel_cif}) and (\ref{eq:kernel_surv}). The 
$Se(c,\tau)$, $Sp_{A}(c,\tau)$ and $Sp_{B}(c,\tau)$ can be estimated by 

\begin{align}
\widehat{Se}(c,\tau) & =\frac{\sum_{i=1}^{n}\widehat{W}_{1i}\cdot1(U_{i}>c)}{\sum_{i=1}^{n}\widehat{W}_{1i}}\nonumber \\
\widehat{Sp}_{A}(c,\tau) & =\frac{\sum_{i=1}^{n}(1-\widehat{W}_{1i})\cdot1(U_{i}\le c)}{\sum_{i=1}^{n}(1-\widehat{W}_{1i})}\label{eq:sen_est}\\
\widehat{Sp}_{B}(c,\tau) & =\frac{\sum_{i=1}^{n}(1-\sum_{k=1}^{K}\widehat{W}_{ki})\cdot1(U_{i}\le c)}{\sum_{i=1}^{n}(1-\sum_{k=1}^{K}\widehat{W}_{ki}).}\nonumber 
\end{align}
The estimator of sensitivity can be justified theoretically
as 

\begin{align*}
Se(c,\tau) & =P(U>c\vert T\le\tau,\delta=1)\\
 & =\frac{E\big(1\{U>c\}\times1\{T\le\tau,\delta=1\}\big)}{E\big(1\{T\le\tau,\delta=1\}\big)}\\
 & =\frac{E\Big\{1\{U>c\}\times E\big(1\{T\le\tau,\delta=1\}\vert\tilde{T},\tilde{\delta},U\big)\Big\}}{E\Big\{ E\big(1\{T\le\tau,\delta=1\}\vert\tilde{T},\tilde{\delta},U\big)\Big\}}\\
 & =\textrm{lim}_{n\rightarrow\infty}\frac{\sum_{i=1}^{n}1(U_{i}>c)\cdot P(T_{i}\le\tau,\delta_{i}=1\vert\tilde{T}_{i},\tilde{\delta}_{i},U_{i})}{\sum_{i=1}^{n}P(T_{i}\le\tau,\delta_{i}=1\vert\tilde{T}_{i},\tilde{\delta}_{i},U_{i}).}
\end{align*}

The justification for the specificity estimator is similar. The time-dependent
ROC curve is an increasing function obtained by plotting the time-dependent
sensitivity and 1-specificity over a range of threshold $c$'s. By definition,
the AUC can be calculated by trapezoidal integration: $\int_{0}^{1}\widehat{ROC}_{A}(x,\tau)dx=\int_{0}^{1}\widehat{Se}(\widehat{Sp}_{A}^{-1}(1-x,\tau),\tau)dx$
and $\int_{0}^{1}\widehat{ROC}_{B}(x,\tau)dx=\int_{0}^{1}\widehat{Se}(\widehat{Sp}_{B}^{-1}(1-x,\tau),\tau)dx$.
Alternatively, it can be estimated by the empirical estimator of the proportion
of concordance pairs, with the proposed weight estimator $\widehat{W}_{1i}$:

\begin{align}
\widehat{AUC}_{A}(\tau) & =\frac{\sum_{i}\sum_{j}\widehat{W}_{1i}(1-\widehat{W}_{1i})\cdot1(U_{i}>U_{j})}{\sum_{i}\sum_{j}\widehat{W}_{1i}(1-\widehat{W}_{1i})}\nonumber \\
\widehat{AUC}_{B}(\tau) & =\frac{\sum_{i}\sum_{j}\widehat{W}_{1i}(1-\sum_{k=1}^{K}\widehat{W}_{ki})\cdot1(U_{i}>U_{j})}{\sum_{i}\sum_{j}\widehat{W}_{1i}(1-\sum_{k=1}^{K}\widehat{W}_{ki}).}\label{eq:AUC_est}
\end{align}
In practice, we can add $0.5\times1(U_{i}=U_{j})$ to the group of
$1(U_{i}>U_{j})$ to account for ties between the $U$'s. The theoretical
justification for the AUC estimators above is as follows.

\begin{align}
AUC_{A}(\tau) & =P(U_{i}>U_{j}\vert T_{i}\le\tau,\delta_{i}=1,\{T_{j}>\tau\}\cup\{T_{j}\le\tau,\delta_{j}\ne1\})\nonumber \\
 & =\frac{E\big(1(T_{i}\le\tau,\delta_{i}=1)\times1(\{T_{j}>\tau\}\cup\{T_{j}\le\tau,\delta_{j}\ne1\})\times1(U_{i}>U_{j})\big)}{E\big(1(T_{i}\le\tau,\delta_{i}=1)\times1(\{T_{j}>\tau\}\cup\{T_{j}\le\tau,\delta_{j}\ne1\})\big)}\nonumber \\
 & =\frac{E\Big\{1(U_{i}>U_{j})\cdot E\big(1(T_{i}\le\tau,\delta_{i}=1)\cdot1(\{T_{j}>\tau\}\cup\{T_{j}\le\tau,\delta_{j}\ne1\})\vert\tilde{T}_{i},\tilde{\delta_{i}},U_{i},\tilde{T_{j}},\tilde{\delta_{j}},U_{j}\big)\Big\}}{E\Big\{ E\big(1(T_{i}\le\tau,\delta_{i}=1)\cdot1(\{T_{j}>\tau\}\cup\{T_{j}\le\tau,\delta_{j}\ne1\})\vert\tilde{T}_{i},\tilde{\delta_{i}},U_{i},\tilde{T_{j}},\tilde{\delta_{j}},U_{j}\big)\Big\}}\nonumber \\
 & =\textrm{lim}_{n\rightarrow\infty}\frac{\sum_{i}\sum_{j}1(U_{i}>U_{j})\cdot P(T_{i}\le\tau,\delta_{i}=1\vert\tilde{T}_{i},\tilde{\delta}_{i},U_{i})\cdot\big(1-P(T_{i}\le\tau,\delta_{i}=1\vert\tilde{T}_{i},\tilde{\delta}_{i},U_{i})\big)}{\sum_{i}\sum_{j}P(T_{i}\le\tau,\delta_{i}=1\vert\tilde{T}_{i},\tilde{\delta}_{i},U_{i})\cdot\big(1-P(T_{i}\le\tau,\delta_{i}=1\vert\tilde{T}_{i},\tilde{\delta}_{i},U_{i})\big)}\nonumber \\
 & =\textrm{lim}_{n\rightarrow\infty}\frac{\sum_{i}\sum_{j}1(U_{i}>U_{j})\times W_{1i}\times\big(1-W_{1i}\big)}{\sum_{i}\sum_{j}W_{1i}\times\big(1-W_{1i}\big)}\label{eq:AUC_justify-1}
\end{align}

A similar justification for $AUC_{B}(\tau)$ is obtained by replacing $\big(1-W_{1i}\big)$ in the
formula (\ref{eq:AUC_justify-1}) with $(1-\sum_{k=1}^{K}W_{ki})$ for the control definition B.
In our numerical studies, the estimator in (\ref{eq:AUC_est}) is
almost identical (up to four digits after the decimal) to the AUC estimator obtained by trapezoidal integration.
The confidence intervals for sensitivity, specificity and AUC can be
estimated numerically by bootstrapping.

\subsection{The Proposed Weighting Estimators for the Brier Score}

By definition, the Brier score is the expected quadratic loss function between the
true disease status $1(T_{i}\le\tau,\delta_{i}=1)$ and the risk score for
event 1, $U_{i}=\pi_{1}(\tau\vert\mathbf{Z}_{i})$, calculated from a prognostic
model to be evaluated. We propose the following estimator for the Brier score, weighting observations according to their probability of having the event of interest:
\[
\widehat{Brier}(\tau)=\dfrac{1}{n}\sum_{i=1}^{n}\Big(\widehat{W}_{1i}\cdot(1-U_{i})^{2}+(1-\widehat{W}_{1i})\cdot(0-U_{i})^{2}\Big).
\]
The justification for consistency of the above estimator is

\begin{align*}
Brier(\tau) & =E\Big\{1(T_{i}\le\tau,\delta_{i}=1)-U_{i}\Big\}^{2}\\
 & =E\Big\{ E\Big(\Big[1(T_{i}\le\tau,\delta_{i}=1)-U_{i}\Big]{}^{2}\vert\tilde{T}_{i},\tilde{\delta}_{i},U_{i}\Big)\Big\}\\
 & =E\Big\{ P(T_{i}\le\tau,\delta_{i}=1\vert\tilde{T}_{i},\tilde{\delta}_{i},U_{i})\cdot(1-U_{i})^{2}+(1-P(T_{i}\le\tau,\delta_{i}=1\vert\tilde{T}_{i},\tilde{\delta}_{i},U_{i}))\cdot(0-U_{i})^{2}\Big\}\\
 & =\textrm{lim}_{n\rightarrow\infty}\frac{1}{n}\sum_{i=1}^{n}\Big(W_{1i}\cdot(1-U_{i})^{2}+(1-W_{1i})\cdot(0-U_{i})^{2}\Big).
\end{align*}

Similarly, the $AbsErr(\tau)$ and $KL(\tau)$ can be estimated with
the proposed conditional probability weight:
\[
\widehat{KL}(\tau)=-\frac{1}{n}\sum_{i=1}^{n}\Big(\widehat{W}_{1i}\text{\ensuremath{\cdot}log}U_{i}+(1-\widehat{W}_{1i})\text{\ensuremath{\cdot}log}(1-U_{i})\Big)
\]
and
\[
\widehat{AbsErr}(\tau)=\frac{1}{n}\sum_{i=1}^{n}\Big(\widehat{W}_{1i}\cdot(1-U_{i})+(1-\widehat{W}_{1i})\cdot U_{i}\Big).
\]
\\
\\
To summarize, the proposed method is a nonparametric
method for estimating the time-dependent predictive accuracy for competing
risk data. It extends the methodology in Li et al. \cite{li2016simple} for a single right-censored time-to-event outcome to competing risk outcomes and to time-dependent calibration metrics. The proposed methodology has a connection to some existing methods. In the context of semi-competing risks with interval censoring, Jacqmin-Gadda et al. \cite{jacqmin2016receiver} proposed an imputation estimator that weights the data with a similar conditional probability of observing an
event in the presence of interval censoring. But their estimator of the conditional probability
is calculated from a parametric illness-death model using the survival and
marker. Schemper \& Henderson \cite{schemper2000predictive} also proposed an imputation method with a Cox model-based estimator for $AbsErr(\tau)$. But this method was shown to be biased when the
prognostic model was misspecified, and an alternative IPCW estimator was proposed in that situation \cite{schmid2011robust}. In contrast, our method is nonparametric, without modeling assumptions, and is applicable to both time-dependent discrimination and calibration metrics. We demonstrated the robustness of the nonparametric method to the selection of tuning parameters in Section 4.3.

\section{Simulation}

In this section, we present simulation studies to evaluate the performance
of the proposed method in estimating both the time-dependent ROC and time-dependent
Brier score in the context of competing risk data. The performance of the proposed
method is compared with those of NNE \cite{saha2010time,zheng2012evaluating}
and IPCW \cite{blanche2013estimating,schoop2011quantifying} methods
from the published literature.

\subsection{Simulation design}

We generate two independent baseline covariates $\boldsymbol{Z}_{i}=(Z_{i1},Z_{i2})$,
where $Z_{i1}$ is a biomarker variable of standard normal distribution,
and $Z_{i2}$ is a baseline characteristic (e.g., gender) of Bernoulli distribution with probability 0.5. The event times are generated according to a Fine-Gray model by using the procedure described in Fine \& Gray \cite{fine1999proportional} with a baseline
sub-distribution hazard (SDH) function and additive covariate effects
on the log SDH. The baseline SDH of event 1 follows a mixture
of Weibull distribution with scale $\lambda_{1}$ and shape $\alpha_{1}$, and a point mass with probability $1-p$ at $\infty$. The log SDH
ratios for covariates $Z_{i1}$ and $Z_{i2}$ are denoted by $\boldsymbol{\beta}=(\beta_{1},\beta_{2})'$ for event 1 and $\boldsymbol{\gamma}=(\gamma_{1},\gamma_{2})'$ for event 2. In our simulations, we set $\boldsymbol{\beta}= (-0.6,0.5)'$,
and $\boldsymbol{\gamma}= (-0.1,-0.2)'$. The event indicator is generated from a Bernoulli distribution with the probability of event 1 being $P_{1}=F_{1}(\infty\vert \boldsymbol{Z})=1-(1-p)^{\textrm{exp}(\boldsymbol{Z \beta}))}$.
The values of $p$ are set to be $(0.22,0.42,0.61)$ to achieve 30\%,
50\% and 70\% of event 1 given the covariate effects. Unless otherwise specified, the random censoring times are generated from a mixture
of uniform distributions on the intervals of $(0,3]\cup(3,6]\cup(6,9]\cup(9,12]\cup(12,15]\cup(15,18]$.
We adjust the probability of falling into each interval to control
the censoring rate. Each simulated data set consists of i.i.d. samples
of $\{(\tilde{T}_{i},U_{i},\tilde{\delta}_{i}),i=1,2,\ldots n\}$:
the observed event time $\tilde{T_{i}}$ is the true event time or censoring
time, whichever comes first; the prognostic score $U_{i}$ is the probability
of experiencing event 1 prior to $\tau$; and the event indicator $\tilde{\delta}_{i}$
takes values of 0, 1, or 2. We use the simulated data sets as validation
data sets to evaluate the predictive accuracy of prognostic score
$U_{i}$ at horizon $\tau$.

We organize the simulation scenarios into a $3\times2\times2$ factorial
design. We consider three proportions for event type 1 (70\%, 50\%
and 30\%), two levels of censoring rates (medium: 25\%-30\% and high:
45\%-50\%) and two sample sizes ($300$ and $600$). The predictive
accuracy is estimated at a time horizon $\tau$, which is approximately
at the 65\% quantile of the observed event time distribution for each
scenario. We compute the true values of $AUC(\tau)$ and $Brier(\tau)$
by a Monte Carlo method using 20,000 independent data sets without
censoring. The prognostic score $U_{i}$ is computed from the true
CIF at $\tau$: $F_{1}(\tau;\boldsymbol{Z})=P(T\le\tau,\delta=1\vert\boldsymbol{Z})=1-\{1-p(1-e^{-\lambda_{1}\tau^{\alpha_{1}}})\}^{\textrm{exp}(\boldsymbol{Z \beta})}$.
In each setting, $500$ Monte Carlo repetitions are performed and
the results are aggregated to compute the bias percentage (bias\%) and MSE in estimating $AUC(\tau)$ and $Brier(\tau)$. 

The results are presented in Sections 4.2, 4.3 and 4.4. In Section
4.2, we compare the finite sample performance of the proposed method
with those of some existing methods. For the estimation of the time-dependent
ROC, we compare the proposed estimator with those of the 
NNE \cite{saha2010time,zheng2012evaluating} and IPCW methods \cite{blanche2013estimating}.
The NNE method is available in the \texttt{R} package \texttt{CompRisksROC} \cite{saha2010time} for
Definition B of Section \ref{subsec:tdROC_comprsk}, and package
\texttt{SurvCompetingRisk} \cite{zheng2012evaluating} for Definition A. The IPCW method is available
in the R package \texttt{timeROC} \cite{blanche2013estimating}.
For the estimation of the Brier score, the proposed estimator is compared
with that of the IPCW method \cite{schoop2011quantifying}. Since the proposed
method is nonparametric with a tuning parameter (bandwidth or span),
we study the sensitivity of the results to the tuning parameter selection in Section 4.3 and compare the performance with that of another
nonparametric method (NNE) that also uses a bandwidth. In Section 4.4,
we take a closer examination of the relative performance of the proposed
method and IPCW when the censoring time is correlated with the risk score. We consider two
versions of IPCW methods that have been reported in the literature. The first one is the IPCW.KM method \cite{blanche2013estimating,graf1999assessment}, where the censoring distribution in the weight function is estimated by the Kaplan-Meier estimator without conditioning on the risk score: 
\begin{equation}
\widehat{W}_{i}^{IPCW.KM}(\tau)=\frac{1(\tilde{T_{i}}\le\tau,\tilde{\delta_{i}}\ne0)}{\widehat{G}(\tilde{T_{i}})}+\frac{1(\tilde{T_{i}}>\tau)}{\widehat{G}(\tau)}.\label{eq:IPCW.KM}
\end{equation}
The second one is the IPCW.Cox method \cite{gerds2006consistent,schoop2011quantifying}, where the censoring distribution in the weight function is estimated from a Cox proportional hazard model, conditioning on the risk score 
\begin{equation}
\widehat{W}_{i}^{IPCW.Cox}(\tau)=\frac{1(\tilde{T_{i}}\le\tau,\tilde{\delta_{i}}\ne0)}{\widehat{G}(\tilde{T_{i}}\vert U)}+\frac{1(\tilde{T_{i}}>\tau)}{\widehat{G}(\tau\vert U)}.\label{eq:IPCW.Cox}
\end{equation}

The sensitivity, specificity and Brier score based on the IPCW weight
$\widehat{W}^{IPCW}(t)=1/\widehat{G}(t\vert \cdot)$ from the equations above are estimated as 
\begin{align*}
\widehat{Se}^{IPCW}(c,\tau) & =\dfrac{\sum_{i=1}^{n}1(U_{i}>c)\times1(\tilde{T}_{i}\le\tau,\tilde{\delta}_{i}=1)\times\widehat{W}_{i}^{IPCW}(\tilde{T}_{i})}{\sum_{i=1}^{n}1(\tilde{T}_{i}\le\tau,\tilde{\delta}_{i}=1)\times\widehat{W}_{i}^{IPCW}(\tilde{T}_{i})}\\\\
\widehat{Sp}_{A}^{IPCW}(c,\tau) & = \dfrac{\sum_{i=1}^{n}1(U_{i}\le c)\times1(\tilde{T}_{i}>\tau)\times\widehat{W}_{i}^{IPCW}(\tau)}{\sum_{i=1}^{n}1(\tilde{T}_{i}>\tau)\times\widehat{W}_{i}^{IPCW}(\tau)}\\\\
\widehat{Sp}_{B}^{IPCW}(c,\tau) & = \dfrac{\sum_{i=1}^{n}1(U_{i}\le c)\times\Big(1(\tilde{T}_{i}>\tau)\cdot\widehat{W}_{i}^{IPCW}(\tau)+1(\tilde{T}_{i}\le\tau,\tilde{\delta}_{i}\notin\{0,1\}\cdot\widehat{W}_{i}^{IPCW}(\tilde{T}_{i})\Big)}{\sum_{i=1}^{n}\Big\{1(\tilde{T}_{i}>\tau)\cdot\widehat{W}_{i}^{IPCW}(\tau)+1(\tilde{T}_{i}\le\tau,\tilde{\delta}_{i}\notin\{0,1\}\cdot\widehat{W}_{i}^{IPCW}(\tilde{T}_{i})\Big\}}\\\\
\widehat{Brier}^{IPCW}(\tau) & =\frac{1}{n}\sum_{i=1}^{n}\Big(1(\tilde{T}_{i}\le\tau,\tilde{\delta}_{i}=1)-\pi(\tau\vert \boldsymbol{Z})\Big)^{2}\times\widehat{W_{i}}^{IPCW}. 
\end{align*}

\subsection{Simulation results on the finite sample performance of the proposed method.}

Table 1 shows the performances of the proposed method, IPCW
and NNE for estimating $\widehat{AUC}_{A}(\tau)$ and $\widehat{AUC}_{B}(\tau)$
under 12 simulation scenarios. For IPCW,
we use the estimator with the weight calculated by  (\ref{eq:IPCW.KM}).
In general, the proposed method has smaller bias than the IPCW, and
the magnitude of the bias is negligible ($<1\% $ in most settings).
The NNE method has notably larger bias, especially for $\widehat{AUC}(\tau)$.
The MSE for the proposed method is also the smallest among the three methods
studied. Table 2 shows the performance of the proposed estimators
and IPCW estimators for estimating the Brier score. The bias percentages
of the proposed estimator are less than 1.5\% in all settings and are
in general smaller than those from the IPCW method. The MSEs of the
proposed estimators are also similar to or smaller than those from the
IPCW method. The NNE method was proposed in the literature only for estimating the AUC and hence was not included in the simulation about the Brier score. We conclude
that the proposed method performs similarly or better than the IPCW
method, and both methods are substantially better than the NNE method.

\subsection{Simulation results on the sensitivity to tuning parameter selection.}

One advantage of the proposed method is that it is nonparametric,
which prevents the predictive accuracy from being affected
by the modeling assumptions involved in calculating the predictive accuracy metrics themselves. However, it
does involve a tuning parameter, which is the bandwidth or span that
is used in the kernel weight calculation. Therefore, it is important
to study whether this estimator is sensitive to the tuning parameter
selection. Since the NNE method also uses the tuning parameter, and
to our knowledge no previous work has studied its sensitivity
to the tuning parameter selection, we include that method in the comparison.
Table 3 presents the performance of the proposed and NNE
methods in estimating the AUC under different spans. This table
only includes the results with 70\% of event
1; the results under other scenarios lead to the same general conclusion
and are hence omitted for brevity. When the $span$ varies from 0.05
to 0.5, the proposed method is quite stable and the bias remains under
1.5\% in all scenarios. Slightly larger biases are observed under
two scenarios: small sample size ($n=300$) with small span $(span=0.05$),
and large sample size $(n=600$) with unrealistically large span ($span=0.5$).
When both the sample size and span are small, there is not enough data
for estimation; and when both the sample size and span are
large, bias may be introduced. In contrast, the NNE estimator
is very sensitive to the span and can result in a large bias when the
span is not chosen properly. We speculate that this led to the relatively
large bias shown in Table 1. A similar performance is observed in Table 4
when the Brier score is estimated. A heuristic explanation of the robustness
of the proposed method to the tuning parameter selection is as follows. First,
the tuning parameter only affects subjects who are censored prior
to time $\tau$ because their disease status at $\tau$ is unknown.
This is a smaller proportion than the overall censoring proportion
of the data. Second, the probability weight $W_{1i}=\dfrac{F_{1}(\tau\vert U_{i})-F_{1}(\tilde{T}_{i}\vert U_{i})}{S(\tilde{T}_{i}\vert U_{i})}$
is defined as the ratio of two conditional probabilities for subjects
censored before $\tau$.
The numerator of $W_{1i}$ can be expressed as the cause-specific
survival probability between $\tilde{T}_{i}$ and $\tau$: $S_{1}(\tilde{T}_{i}\vert U_{i})-S_{1}(\tau\vert U_{i})=Pr(\tilde{T}_{i}<T\le\tau,\delta_{i}=1\vert U_{i})$;
and the denumerator is the overall survival probability beyond $\tilde{T}_{i}$. The asymptotic bias of two conditional survival probabilities as
a function of bandwidth are in the same direction \cite{bordes2011uniform}.
Therefore, the bias of their ratio can be canceled out to some extent,
particularly when $\tilde{T}_{i}$ and $\tau$ are close.

\subsection{Simulation results for the performance of the proposed method under dependent
censoring. }

In this section, we compare the proposed method and IPCW under a
dependent censoring scenario where the event time $T$ and censoring
time $C$ are marginally dependent but are conditionally independent
given the risk score $U$. In practice, the censoring time is often correlated with baseline covariates. Since $U$ is a function of these covariates,
$C$ and $U$ may also be correlated. Literature on the time-dependent
ROC and time-dependent Brier score describes estimation under dependent censoring of this kind using the
IPCW approach, where a Cox model is used to estimate the censoring distribution, conditioning on the risk score \cite{gerds2006consistent,schoop2011quantifying}. In contrast,
our proposed method does not model the censoring distribution, which is a nuisance for scientific purposes. We directly estimate the conditional survival and CIF nonparametrically. In this simulation, we consider two settings. In setting $(a)$, we generate censoring
time $C_{i}$ from a Weibull$(\lambda_{c},\alpha_{c})$ distribution
with the mean $\mu_{C}=\frac{\Gamma(1+1/\alpha_{c})}{\lambda_{c}^{1/\alpha_{c}}}=a*1\{(\zeta >0.4)\cup(\zeta <-0.6)\}+b*1\{-0.6\le \zeta \le0.4\}$, where $\zeta=\boldsymbol{Z \beta}$ is a monotone transformation of $U$. 
Different values of $(a,b)$ and $\alpha_{c}$ are chosen to achieve
a medium or high censoring rate. The dependency between the censoring
distribution and $U$ is not monotone and cannot be correctly estimated
by a proportional hazard model. We use setting (a) to study the robustness of the methods to model misspecification. In setting $(b)$, we generate
the censoring time from a Cox model on $\zeta$, so that the censoring time is correctly modeled by the IPCW. For both
settings, we compare the performance of the proposed method and IPCW methods with both weight estimators (\ref{eq:IPCW.KM}) and (\ref{eq:IPCW.Cox}).

Tables 5 and 7 compare the performance of the proposed method with that of the IPCW
in estimating $AUC(\tau)$. All bias percentages for the proposed
method are under 1.5\% and 1\% for settings $(a)$ and $(b)$, respectively.
In contrast, the IPCW.KM method, which ignores the dependent censoring,
produces results with a large bias under both mechanisms. Compared to
IPCW.KM, the IPCW.Cox estimator in setting $(a)$ alleviates the bias
by accounting for the dependence but still has larger bias and MSE
than the proposed method, especially when the type 1 event rate is
low (e.g., 30\%). When the censoring times are generated from the
Cox model in setting $(b)$, the bias from the IPCW.Cox method is controlled
under 1.5\% but is still slightly larger than that from the proposed method
in general. This indicates that the proposed method is more robust
than the IPCW methods under different dependence structures of $C$ and
$U$. 

Tables 6 and 8 present similar comparisons between the proposed method
and IPCW in estimating the Brier score. The overall performance
is similar to that of $\widehat{AUC}(\tau)$. However, we notice that when the IPCW.Cox method is used under a misspecified censoring mechanism in setting $(a)$, it produces a larger bias in the estimation of the Brier score than the AUC. In contrast, the performance of IPCW.Cox under
setting $(b)$ is similar in both estimands, with the biases well
controlled under 1.5\%. The results indicate that estimation of $\widehat{Brier}(\tau)$
appears to be more sensitive to misspecification than that of $\widehat{AUC}(\tau)$.
We speculate that this is because $AUC(\tau)$ is based on the rankings
of the data, whereas $Brier(\tau)$ measures the actual deviation
from the true status in quantity and therefore is more sensitive to the
misspecification of the estimation procedure.

The results above suggest that our nonparametric method does not suffer from
bias caused by model dependence. The rationale for developing a nonparametric
estimation method is that the estimator of a predictive accuracy metric
should be an objective reflection of the model under evaluation, without
introducing another source of bias due to the modeling assumption of
the estimation method. In this spirit, one can extend
the IPCW method by using a nonparametric estimator
for the conditional distribution of the censoring time given the risk score. But from a clinical perspective, this conditional distribution is less intuitive than directly
modeling the conditional survival distribution, which offers additional
insight into the relationship between the risk score and disease development. In addition, the relationship between the risk score and the survival time is expected to be monotone by the definition of the ROC, but this is not necessarily the case for the relationship between the risk score and the censoring time. The nonparametric smoothing literature suggests that the nonparametric regression result is less sensitive to the tuning parameters when the relationship between the outcome and covariate is monotone \cite{meyer2008inference}.  

In summary, the simulation results from Table 1 to Table 8 demonstrate
that the proposed method has similar or better performance than other
published methods. While the NNE method only estimates the time-dependent
ROC, the proposed method works with the time-dependent ROC, time-dependent
Brier score and other predictive accuracy metrics, with notably smaller
bias and MSE. Unlike the NNE, the proposed method is
robust to tuning parameter selection, which makes it easy to use in
practice. As a nonparametric method, the proposed method outperforms
the IPCW under dependent censoring, particularly in light of the
possibility that IPCW may use a misspecified model for the censoring
distribution.

\section{Application}

We illustrate the proposed method with a data set from AASK, a randomized
clinical trial for 1,094 patients with chronic kidney disease, whose
baseline estimated glomerular filtration rates (eGFRs) were between $20-65\ \textrm{mL/min/1.73m}^{2}$
\cite{wright2002effect}. The patients were followed for 6.5 years
during the trial period. Among them, 179 developed ESRD and 85 died before developing ESRD. We evaluate the predictive
accuracy of a prognostic risk score developed from a proportional
sub-distributional hazard model with five baseline covariates: the
eGFR, urine protein creatinine
ratio, age, gender, the randomized blood pressure group (low and medium) and the randomized anti-hypertensive therapy (ramipril, metoprolol, amlodipine). The prognostic
score is the predicted CIF for ESRD at prespecified horizons.

Figure 1 compares the time-dependent ROC curves estimated from the
proposed method (red), IPCW.KM (black), IPCW.Cox (blue) and NNE (green) 
at three predictive horizons: 3, 4 and 5 years from baseline. The span used in the proposed and NNE methods is 0.05, which includes 5\% of the neighborhood data. The two rows in the panel present the estimated ROC curves based on the two definitions (Section 2.2).
Definition A discriminates patients with ESRD within $\tau$ years from ESRD-free patients, which include patients who are event-free and who die by year $\tau$.
Definition B discriminates patients with ESRD within $\tau$ years from those who are event-free at year $\tau$. 
The ROC curves from the two IPCW methods, IPCW.KM and IPCW.Cox, are
almost identical. The curves by IPCW and the proposed method are also very
close, and the differences between the $\widehat{AUC}(\tau)$ are within
5\%. The estimated $\widehat{AUC}^{A}(\tau)$ and $\widehat{AUC}^{B}(\tau)$ are also very close within the different estimation methods except
for NNE. This indicates that the sub-distribution hazard
model we used can discriminate well between ESRD patients and ESRD-free or event-free patients. A possible explanation is that the patients who died in the study period are a relatively small population and may have died from causes unrelated to kidney disease. Therefore, adding these patients to the control group may not substantially change the discrimination
of the risk score, which primarily consists of risk factors for ESRD. There
is some discussion of how to use different definitions of controls in the ROC estimation \cite{zheng2012evaluating}; the choice is related to the clinical context and here we provide estimation methods
for both. 

In Figure 2, we show further results of our study of the proposed and NNE methods with varying spans of
0.05, 0.1, and 0.3. The proposed
method produces stable $\widehat{AUC}(\tau)$ around 0.88 while the NNE
method is very sensitive to the span specification. This result is consistent
with the simulation results in Table 4. Such robustness to the tuning parameter
selection is a very attractive feature for our nonparametric estimator.

The Brier scores over all the predictive horizons are plotted
in Figure 3, along with the percentages of ESRD and censoring at each predictive horizon. The prediction error increases with the predictive horizon. This result implies that the predictive accuracy decreases as the predictive horizon moves away from the time of prediction. Overall the estimated Brier scores are small, between 0 and 0.11. Prior to year 3.5, when there is little censoring,
the three estimation methods produce almost identical results.
When the percentage of censoring increases beyond 3.5 years, the results from the three
methods begin to diverge but the absolute differences among them remain 
small.

\section{Discussion }
In this paper, we propose an analytical framework for estimating
time-dependent predictive accuracy metrics with competing risk data
that are subject to right censoring. The method is illustrated with
the time-dependent ROC and time-dependent Brier score. The
proposed framework first computes a nonparametric estimator of the
conditional probability of the true event status given the observed
data and then uses it to weigh the data in an empirical calculation
of the time-dependent metrics. This is a unified
approach to estimating
the time-dependent ROC, time-dependent Brier score, and time-dependent
metrics constructed from other loss functions. The proposed method requires
no parametric assumptions about the marginal, conditional or joint
distribution of the risk score and time to the event of interest. It can be applied to evaluate the discrimination for a
single biomarker or a risk score constructed from a prognostic model with multiple biomarkers, and to evaluate the
calibration of the prognostic model.
The method is applicable when the censoring time and the risk score are correlated. It is  also insensitive to the tuning parameter specification. Such robustness to the tuning
parameter specification has not been studied in nonparametric estimations of time-dependent predictive accuracy metrics \cite{heagerty2000time,saha2010time,zheng2012evaluating}
and no guidelines are yet available for practical users. When compared with competing methods in simulations, our proposed method demonstrates better
overall performance and robustness to tuning parameters, particularly when the censoring is correlated with the risk score. The R code that implements the proposed methodology is available upon request and will be added to the \texttt{tdROC} package in \texttt{R}.

One limitation with the proposed method is that, like many other nonparametric
methods, it works better with larger sample sizes. When the sample size is very
small, there may not be enough subjects with events for calculating $\widehat{F}_{1}(t\vert U_{i})$ and $\widehat{S}_{T}(t\vert U_{i})$
within some local neighborhoods defined by the kernel. In such case, the bandwidth may need to be increased for those neighborhoods.

\section{Acknowledgment }

The authors declare no potential conflicts of interest with respect
to the research, authorship and publication of this article. This
research was supported by the U.S. National Institutes of Health (grants
5P30CA016672 and 5U01DK103225).

{55}

\newpage{}

\begin{table}[ph]
\begin{centering}
\caption{Simulation results of $\widehat{AUC}_{A}(\tau)$ and $\widehat{AUC}_{B}(\tau)$
for the proposed method, IPCW, and NNE under different event 1 rate (70\%,
50\% and 30\%), censoring rate (Medium: 25-30\%, High: 45-50\%).}
\par\end{centering}
\centering{}%
\begin{tabular}{ccccccccccc}
\hline 
\multirow{2}{*}{Event 1} & \multirow{2}{*}{Censoring} & \multirow{2}{*}{True} & \multirow{2}{*}{n} & \multicolumn{3}{c}{Bias\% $\widehat{AUC}_{A}(\tau)$} &  & \multicolumn{3}{c}{MSE$\times10^{-3}$$\widehat{AUC}_{A}(\tau)$}\tabularnewline
\cline{5-7} \cline{9-11} 
 &  &  &  & {\small{}proposed} & {\small{}IPCW} & {\small{}NNE} &  & {\small{}proposed} & {\small{}IPCW} & {\small{}NNE}\tabularnewline
\hline 
\multirow{4}{*}{70\%} & \multirow{2}{*}{Medium} & \multirow{2}{*}{0.698} & 300 & 0.358 & 0.720 & -3.024 &  & 1.281 & 1.299 & 1.600\tabularnewline
 &  &  & 600 & 0.789 & 1.129 & -1.577 &  & 0.591 & 0.638 & 0.646\tabularnewline
\cline{2-11} 
 & \multirow{2}{*}{High} & \multirow{2}{*}{0.693} & 300 & 0.487 & 0.950 & -2.986 &  & 1.605 & 1.620 & 1.853\tabularnewline
 &  &  & 600 & 0.700 & 1.158 & -1.714 &  & 0.797 & 0.846 & 0.869\tabularnewline
\hline 
\multirow{4}{*}{50\%} & \multirow{2}{*}{Medium} & \multirow{2}{*}{0.691} & 300 & 0.519 & 0.877 & -2.869 &  & 1.567 & 1.593 & 1.826\tabularnewline
 &  &  & 600 & 0.447 & 0.800 & -2.046 &  & 0.774 & 0.789 & 0.931\tabularnewline
\cline{2-11} 
 & \multirow{2}{*}{High} & \multirow{2}{*}{0.685} & 300 & 0.106 & 0.651 & -3.430 &  & 1.850 & 1.886 & 2.255\tabularnewline
 &  &  & 600 & 0.827 & 1.311 & -1.718 &  & 0.974 & 1.054 & 1.030\tabularnewline
\hline 
\multirow{4}{*}{30\%} & \multirow{2}{*}{Medium} & \multirow{2}{*}{0.685} & 300 & 0.544 & 0.924 & -2.907 &  & 2.020 & 2.050 & 2.290\tabularnewline
 &  &  & 600 & 0.885 & 1.272 & -1.662 &  & 0.985 & 1.020 & 1.042\tabularnewline
\cline{2-11} 
 & \multirow{2}{*}{High} & \multirow{2}{*}{0.683} & 300 & 0.580 & 0.981 & -2.949 &  & 2.305 & 2.436 & 2.560\tabularnewline
 &  &  & 600 & 0.246 & 0.759 & -2.359 &  & 1.173 & 1.253 & 1.375\tabularnewline
\hline 
 &  &  &  &  &  &  &  &  &  & \tabularnewline
 &  &  &  & \multicolumn{3}{c}{Bias\% $\widehat{AUC}_{B}(\tau)$} &  & \multicolumn{3}{c}{MSE$\times10^{-3}$ $\widehat{AUC}_{B}(\tau)$}\tabularnewline
\cline{5-7} \cline{9-11} 
 &  &  &  & proposed & IPCW & NNE &  & proposed & IPCW & NNE\tabularnewline
\hline 
\multirow{4}{*}{70\%} & \multirow{2}{*}{Medium} & \multirow{2}{*}{0.661} & 300 & 0.500 & 0.887 & -1.027 &  & 1.656 & 1.721 & 1.710\tabularnewline
 &  &  & 600 & 0.985 & 1.347 & -0.796 &  & 0.733 & 0.797 & 0.757\tabularnewline
\cline{2-11} 
 & \multirow{2}{*}{High} & \multirow{2}{*}{0.663} & 300 & 0.585 & 1.054 & -1.270 &  & 1.848 & 1.890 & 1.975\tabularnewline
 &  &  & 600 & 0.693 & 1.164 & -1.336 &  & 0.904 & 0.982 & 1.017\tabularnewline
\hline 
\multirow{4}{*}{50\%} & \multirow{2}{*}{Medium} & \multirow{2}{*}{0.652} & 300 & 0.711 & 1.065 & -1.354 &  & 2.080 & 2.154 & 2.207\tabularnewline
 &  &  & 600 & 0.650 & 1.020 & -1.709 &  & 0.934 & 0.956 & 1.063\tabularnewline
\cline{2-11} 
 & \multirow{2}{*}{High} & \multirow{2}{*}{0.653} & 300 & 0.403 & 0.983 & -1.820 &  & 2.093 & 2.208 & 2.355\tabularnewline
 &  &  & 600 & 1.051 & 1.519 & -1.664 &  & 1.136 & 1.264 & 1.203\tabularnewline
\hline 
\multirow{4}{*}{30\%} & \multirow{2}{*}{Medium} & \multirow{2}{*}{0.672} & 300 & 0.914 & 1.330 & -1.375 &  & 2.404 & 2.487 & 2.492\tabularnewline
 &  &  & 600 & 1.122 & 1.508 & -2.020 &  & 1.162 & 1.229 & 1.325\tabularnewline
\cline{2-11} 
 & \multirow{2}{*}{High} & \multirow{2}{*}{0.672} & 300 & 0.728 & 1.081 & -1.853 &  & 2.632 & 2.828 & 2.726\tabularnewline
 &  &  & 600 & 0.343 & 0.841 & -2.981 &  & 1.294 & 1.418 & 1.619\tabularnewline
\hline 
\end{tabular}
\end{table}

\begin{table}[ph]
\centering{}\caption{Simulation results of $\widehat{Brier}(\tau)$ for the proposed method and IPCW
under different event 1 rate (70\%, 50\% and 30\%), censoring rate
(Medium: 25-30\%, High: 45-50\%).}
\begin{tabular}{ccccccccc}
\hline 
\multirow{2}{*}{Event 1} & \multirow{2}{*}{Censoring} & \multirow{2}{*}{True} & \multirow{2}{*}{n} & \multicolumn{2}{c}{Bias\% BS} &  & \multicolumn{2}{c}{MSE$\times10^{-3}$ BS}\tabularnewline
\cline{5-6} \cline{8-9} 
 &  &  &  & proposed & IPCW.KM &  & proposed & IPCW.KM\tabularnewline
\hline 
\multirow{4}{*}{70\%} & \multirow{2}{*}{Medium} & \multirow{2}{*}{0.195} & 300 & -0.282 & -0.702 &  & 0.143 & 0.146\tabularnewline
 &  &  & 600 & -0.759 & -1.105 &  & 0.066 & 0.070\tabularnewline
\cline{2-9} 
 & \multirow{2}{*}{High} & \multirow{2}{*}{0.182} & 300 & -0.125 & -0.530 &  & 0.179 & 0.176\tabularnewline
 &  &  & 600 & 0.178 & -0.223 &  & 0.096 & 0.097\tabularnewline
\hline 
\multirow{4}{*}{50\%} & \multirow{2}{*}{Medium} & \multirow{2}{*}{0.165} & 300 & -1.337 & -1.575 &  & 0.187 & 0.187\tabularnewline
 &  &  & 600 & -1.478 & -1.698 &  & 0.102 & 0.105\tabularnewline
\cline{2-9} 
 & \multirow{2}{*}{High} & \multirow{2}{*}{0.149} & 300 & -0.697 & -1.052 &  & 0.227 & 0.230\tabularnewline
 &  &  & 600 & -1.137 & -1.401 &  & 0.116 & 0.121\tabularnewline
\hline 
\multirow{4}{*}{30\%} & \multirow{2}{*}{Medium} & \multirow{2}{*}{0.140} & 300 & -1.069 & -1.085 &  & 0.213 & 0.211\tabularnewline
 &  &  & 600 & -0.906 & -0.977 &  & 0.100 & 0.101\tabularnewline
\cline{2-9} 
 & \multirow{2}{*}{High} & \multirow{2}{*}{0.122} & 300 & 0.079 & -0.006 &  & 0.235 & 0.237\tabularnewline
 &  &  & 600 & -0.103 & -0.201 &  & 0.112 & 0.112\tabularnewline
\hline 
\end{tabular}
\end{table}

\begin{table}[ph]
\caption{Simulation results of $\widehat{AUC}_{A}(\tau)$ and $\widehat{AUC}_{B}(\tau)$
for the proposed method and NNE on the robustness of span specification.
Setting: 70\% event 1 rate, censoring rate (Medium: 25-30\%, High:
45-50\%).}

\centering{}%
\begin{tabular}{ccccccccc}
\hline 
\multirow{2}{*}{Censoring} & \multirow{2}{*}{True} & \multirow{2}{*}{n} & \multirow{2}{*}{Span} & \multicolumn{2}{c}{Bias\% $\widehat{AUC}_{A}(\tau)$} &  & \multicolumn{2}{c}{MSE$\times10^{-3}$ $\widehat{AUC}_{A}(\tau)$}\tabularnewline
\cline{5-6} \cline{8-9} 
 &  &  &  & proposed & NNE &  & proposed & NNE\tabularnewline
\hline 
\multirow{8}{*}{Medium} & \multirow{8}{*}{0.698} & \multirow{4}{*}{300} & 0.05 & 1.089 & -2.775 &  & 1.164 & 1.343\tabularnewline
 &  &  & 0.1 & 0.798 & -6.821 &  & 1.342 & 3.141\tabularnewline
 &  &  & 0.3 & -0.202 & -21.348 &  & 1.131 & 22.313\tabularnewline
 &  &  & 0.5 & -0.523 & -25.287 &  & 1.009 & 31.188\tabularnewline
\cline{3-9} 
 &  & \multirow{4}{*}{600} & 0.05 & 0.625 & -2.518 &  & 0.697 & 0.912\tabularnewline
 &  &  & 0.1 & 0.543 & -6.790 &  & 0.606 & 2.649\tabularnewline
 &  &  & 0.3 & -0.278 & -21.502 &  & 0.604 & 22.588\tabularnewline
 &  &  & 0.5 & -0.530 & -25.392 &  & 0.551 & 31.441\tabularnewline
\hline 
\multirow{8}{*}{High} & \multirow{8}{*}{0.693} & \multirow{4}{*}{300} & 0.05 & 1.128 & -3.017 &  & 1.543 & 1.689\tabularnewline
 &  &  & 0.1 & 0.550 & -7.950 &  & 1.517 & 3.944\tabularnewline
 &  &  & 0.3 & -0.355 & -21.845 &  & 1.478 & 23.002\tabularnewline
 &  &  & 0.5 & -1.368 & -25.278 &  & 1.415 & 30.690\tabularnewline
\cline{3-9} 
 &  & \multirow{4}{*}{600} & 0.05 & 0.754 & -2.796 &  & 0.862 & 1.099\tabularnewline
 &  &  & 0.1 & 0.756 & -7.537 &  & 0.703 & 3.143\tabularnewline
 &  &  & 0.3 & -0.355 & -21.991 &  & 0.781 & 23.262\tabularnewline
 &  &  & 0.5 & -1.371 & -25.383 &  & 0.723 & 30.938\tabularnewline
\hline 
 &  &  &  &  &  &  &  & \tabularnewline
 &  &  &  & \multicolumn{2}{c}{Bias\% $\widehat{AUC}_{B}(\tau)$} &  & \multicolumn{2}{c}{MSE$\times10^{-3}$$\widehat{AUC}_{B}(\tau)$}\tabularnewline
\cline{5-6} \cline{8-9} 
 &  &  &  & proposed & NNE &  & proposed & NNE\tabularnewline
\hline 
\multirow{8}{*}{Medium} & \multirow{8}{*}{0.661} & \multirow{4}{*}{300} & 0.05 & 1.222 & 0.705 &  & 1.407 & 1.440\tabularnewline
 &  &  & 0.1 & 0.790 & -0.718 &  & 1.641 & 1.717\tabularnewline
 &  &  & 0.3 & -0.161 & -4.624 &  & 1.358 & 2.079\tabularnewline
 &  &  & 0.5 & -0.247 & -6.180 &  & 1.197 & 2.528\tabularnewline
\cline{3-9} 
 &  & \multirow{4}{*}{600} & 0.05 & 0.810 & 0.144 &  & 0.819 & 0.855\tabularnewline
 &  &  & 0.1 & 0.620 & -0.956 &  & 0.736 & 0.790\tabularnewline
 &  &  & 0.3 & -0.073 & -5.270 &  & 0.754 & 1.814\tabularnewline
 &  &  & 0.5 & -0.202 & -6.651 &  & 0.655 & 2.368\tabularnewline
\hline 
\multirow{8}{*}{High} & \multirow{8}{*}{0.663} & \multirow{4}{*}{300} & 0.05 & 1.158 & 0.534 &  & 1.914 & 1.998\tabularnewline
 &  &  & 0.1 & 0.527 & -1.291 &  & 1.711 & 1.836\tabularnewline
 &  &  & 0.3 & -0.267 & -5.147 &  & 1.671 & 2.528\tabularnewline
 &  &  & 0.5 & -1.199 & -7.061 &  & 1.550 & 3.260\tabularnewline
\cline{3-9} 
 &  & \multirow{4}{*}{600} & 0.05 & 0.796 & -0.092 &  & 0.953 & 0.947\tabularnewline
 &  &  & 0.1 & 0.786 & -1.296 &  & 0.816 & 0.956\tabularnewline
 &  &  & 0.3 & -0.286 & -6.143 &  & 0.904 & 2.336\tabularnewline
 &  &  & 0.5 & -1.231 & -7.893 &  & 0.800 & 3.271\tabularnewline
\hline 
\end{tabular}
\end{table}

\begin{table}[ph]
\begin{centering}
\caption{Simulation results of $\widehat{Brier}(\tau)$ for the proposed method on
the robustness of span specification. Setting: 70\% event 1 rate,
censoring rate (Medium: 25-30\%, High: 45-50\%).}
\par\end{centering}
\centering{}%
\begin{tabular}{cccccc}
\hline 
Censoring & True & n & Span & Bias\% $\widehat{Brier}(\tau)$ & MSE$\times10^{-3}$ $\widehat{Brier}(\tau)$\tabularnewline
\hline 
\multirow{8}{*}{Medium} & \multirow{8}{*}{0.195} & \multirow{4}{*}{300} & 0.05 & -1.159 & 0.143\tabularnewline
 &  &  & 0.1 & -0.470 & 0.162\tabularnewline
 &  &  & 0.3 & 0.065 & 0.121\tabularnewline
 &  &  & 0.5 & 0.614 & 0.140\tabularnewline
\cline{3-6} 
 &  & \multirow{4}{*}{600} & 0.05 & -0.831 & 0.078\tabularnewline
 &  &  & 0.1 & -0.466 & 0.069\tabularnewline
 &  &  & 0.3 & 0.320 & 0.073\tabularnewline
 &  &  & 0.5 & 0.439 & 0.071\tabularnewline
\hline 
\multirow{8}{*}{High} & \multirow{8}{*}{0.182} & \multirow{4}{*}{300} & 0.05 & -1.004 & 0.182\tabularnewline
 &  &  & 0.1 & -0.232 & 0.182\tabularnewline
 &  &  & 0.3 & 0.360 & 0.194\tabularnewline
 &  &  & 0.5 & 1.086 & 0.180\tabularnewline
\cline{3-6} 
 &  & \multirow{4}{*}{600} & 0.05 & -0.439 & 0.099\tabularnewline
 &  &  & 0.1 & -0.069 & 0.079\tabularnewline
 &  &  & 0.3 & 0.425 & 0.096\tabularnewline
 &  &  & 0.5 & 1.086 & 0.095\tabularnewline
\hline 
\end{tabular}
\end{table}

\begin{sidewaystable}
\centering{}\caption{Simulation results of $\widehat{AUC}_{A}(\tau)$ and $\widehat{AUC}_{B}(\tau)$
for the proposed method and IPCW methods under dependent censoring setting
$(a)$. Setting: event 1 rate (70\%, 50\% and 30\%), censoring rate
(Medium: 25-30\%, High: 45-50\%).}
\begin{tabular}{ccccccccccc}
\hline 
\multirow{2}{*}{Event 1} & \multirow{2}{*}{Censoring} & \multirow{2}{*}{True} & \multirow{2}{*}{n} & \multicolumn{3}{c}{Bias\% $\widehat{AUC}_{A}(\tau)$} &  & \multicolumn{3}{c}{MSE$\times10^{-3}$ $\widehat{AUC}_{A}(\tau)$}\tabularnewline
\cline{5-7} \cline{9-11} 
 &  &  &  & proposed & IPCW.KM. & IPCW.Cox &  & proposed & IPCW.KM & IPCW.Cox\tabularnewline
\hline 
\multirow{4}{*}{70\%} & \multirow{2}{*}{Medium} & \multirow{2}{*}{0.698} & 300 & 0.980 & -5.587 & 1.600 &  & 1.359 & 3.042 & 1.308\tabularnewline
 &  &  & 600 & 0.510 & -5.810 & 1.193 &  & 0.588 & 2.300 & 0.586\tabularnewline
\cline{2-11} 
 & \multirow{2}{*}{High} & \multirow{2}{*}{0.692} & 300 & 0.838 & -9.515 & 0.975 &  & 1.718 & 6.109 & 1.463\tabularnewline
 &  &  & 600 & 0.927 & -9.175 & 1.098 &  & 0.864 & 4.910 & 0.740\tabularnewline
\hline 
\multirow{4}{*}{50\%} & \multirow{2}{*}{Medium} & \multirow{2}{*}{0.702} & 300 & -0.673 & 3.351 & 0.322 &  & 1.413 & 1.884 & 2.014\tabularnewline
 &  &  & 600 & -0.720 & 3.264 & 0.490 &  & 0.739 & 1.193 & 1.126\tabularnewline
\cline{2-11} 
 & \multirow{2}{*}{High} & \multirow{2}{*}{0.687} & 300 & 1.312 & -6.183 & 1.073 &  & 2.090 & 3.676 & 1.746\tabularnewline
 &  &  & 600 & 1.332 & -6.192 & 0.995 &  & 1.076 & 2.811 & 0.892\tabularnewline
\hline 
\multirow{4}{*}{30\%} & \multirow{2}{*}{Medium} & \multirow{2}{*}{0.685} & 300 & 0.906 & -4.493 & 1.821 &  & 1.897 & 2.758 & 1.763\tabularnewline
 &  &  & 600 & 0.745 & -4.512 & 1.748 &  & 0.854 & 1.791 & 0.863\tabularnewline
\cline{2-11} 
 & \multirow{2}{*}{High} & \multirow{2}{*}{0.686} & 300 & -0.602 & 4.627 & 1.951 &  & 2.352 & 3.287 & 5.925\tabularnewline
 &  &  & 600 & -0.217 & 4.915 & 2.057 &  & 1.237 & 2.310 & 6.800\tabularnewline
\hline 
 &  &  &  &  &  &  &  &  &  & \tabularnewline
 &  &  &  & \multicolumn{3}{c}{Bias\% $\widehat{AUC}_{B}(\tau)$} &  & \multicolumn{3}{c}{MSE$\times10^{-3}$ $\widehat{AUC}_{B}(\tau)$}\tabularnewline
\cline{5-7} \cline{9-11} 
 &  &  &  & proposed & IPCW.KM. & IPCW.Cox &  & proposed & IPCW.KM & IPCW.Cox\tabularnewline
\hline 
\multirow{4}{*}{70\%} & \multirow{2}{*}{Medium} & \multirow{2}{*}{0.661} & 300 & 1.041 & -8.929 & 1.443 &  & 1.684 & 5.483 & 1.609\tabularnewline
 &  &  & 600 & 0.561 & -9.103 & 0.996 &  & 0.752 & 4.533 & 0.717\tabularnewline
\cline{2-11} 
 & \multirow{2}{*}{High} & \multirow{2}{*}{0.665} & 300 & 0.831 & -13.040 & 0.350 &  & 1.992 & 9.687 & 1.705\tabularnewline
 &  &  & 600 & 0.820 & -12.736 & 0.409 &  & 1.002 & 8.257 & 0.835\tabularnewline
\hline 
\multirow{4}{*}{50\%} & \multirow{2}{*}{Medium} & \multirow{2}{*}{0.674} & 300 & -0.784 & 6.233 & 0.596 &  & 1.809 & 3.367 & 2.972\tabularnewline
 &  &  & 600 & -0.536 & 6.322 & 0.876 &  & 0.874 & 2.601 & 1.622\tabularnewline
\cline{2-11} 
 & \multirow{2}{*}{High} & \multirow{2}{*}{0.663} & 300 & 1.373 & -9.341 & 0.484 &  & 2.368 & 6.113 & 1.981\tabularnewline
 &  &  & 600 & 1.559 & -9.275 & 0.472 &  & 1.222 & 5.000 & 0.994\tabularnewline
\hline 
\multirow{4}{*}{30\%} & \multirow{2}{*}{Medium} & \multirow{2}{*}{0.670} & 300 & 1.096 & -8.818 & 2.223 &  & 2.356 & 5.948 & 2.141\tabularnewline
 &  &  & 600 & 0.948 & -8.836 & 2.098 &  & 1.011 & 4.679 & 1.028\tabularnewline
\cline{2-11} 
 & \multirow{2}{*}{High} & \multirow{2}{*}{0.676} & 300 & -0.266 & 7.490 & 2.504 &  & 2.648 & 5.034 & 7.830\tabularnewline
 &  &  & 600 & -0.113 & 7.651 & 2.043 &  & 1.329 & 3.928 & 8.537\tabularnewline
\hline 
\end{tabular}
\end{sidewaystable}

\begin{sidewaystable}
\caption{Simulation results of $\widehat{Brier}(\tau)$ for the proposed method and
IPCW methods under dependent censoring setting $(a)$. Setting: event
1 rate (70\%, 50\% and 30\%), censoring rate (Medium: 25-30\%, High:
45-50\%).}

\centering{}%
\begin{tabular}{ccccccccccc}
\hline 
\multirow{2}{*}{Event 1} & \multirow{2}{*}{Censoring} & \multirow{2}{*}{True} & \multirow{2}{*}{n} & \multicolumn{3}{c}{Bias\% BS} &  & \multicolumn{3}{c}{MSE$\times10^{-3}$ BS}\tabularnewline
\cline{5-7} \cline{9-11} 
 &  &  &  & proposed & IPCW.KM & IPCW.Cox &  & proposed & IPCW.KM & IPCW.Cox\tabularnewline
\hline 
\multirow{4}{*}{70\%} & \multirow{2}{*}{Medium} & \multirow{2}{*}{0.195} & 300 & -0.849 & 6.737 & 1.533 &  & 0.160 & 0.318 & 0.152\tabularnewline
 &  &  & 600 & -0.625 & 6.823 & 1.729 &  & 0.082 & 0.249 & 0.082\tabularnewline
\cline{2-11} 
 & \multirow{2}{*}{High} & \multirow{2}{*}{0.175} & 300 & -0.576 & 11.201 & 3.742 &  & 0.272 & 0.570 & 0.242\tabularnewline
 &  &  & 600 & -0.494 & 11.240 & 3.776 &  & 0.128 & 0.478 & 0.137\tabularnewline
\hline 
\multirow{4}{*}{50\%} & \multirow{2}{*}{Medium} & \multirow{2}{*}{0.187} & 300 & -0.072 & -5.733 & -1.541 &  & 0.155 & 0.289 & 0.209\tabularnewline
 &  &  & 600 & -0.098 & -5.775 & -1.709 &  & 0.080 & 0.203 & 0.108\tabularnewline
\cline{2-11} 
 & \multirow{2}{*}{High} & \multirow{2}{*}{0.153} & 300 & -0.768 & 8.387 & 2.426 &  & 0.286 & 0.390 & 0.244\tabularnewline
 &  &  & 600 & -1.277 & 8.219 & 2.280 &  & 0.126 & 0.258 & 0.110\tabularnewline
\hline 
\multirow{4}{*}{30\%} & \multirow{2}{*}{Medium} & \multirow{2}{*}{0.142} & 300 & -0.584 & 5.406 & 1.359 &  & 0.222 & 0.258 & 0.206\tabularnewline
 &  &  & 600 & -0.892 & 4.900 & 0.881 &  & 0.097 & 0.134 & 0.089\tabularnewline
\cline{2-11} 
 & \multirow{2}{*}{High} & \multirow{2}{*}{0.130} & 300 & -0.081 & -8.491 & -2.294 &  & 0.236 & 0.343 & 0.252\tabularnewline
 &  &  & 600 & -0.555 & -8.617 & -2.244 &  & 0.109 & 0.233 & 0.127\tabularnewline
\hline 
\end{tabular}
\end{sidewaystable}

\begin{sidewaystable}
\centering{}
\caption{Simulation results of $\widehat{AUC}_{A}(\tau)$ and $\widehat{AUC}_{B}(\tau)$
for the proposed method and IPCW methods under dependent censoring setting
$(b)$. Setting: event 1 rate (70\%, 50\% and 30\%), censoring rate
(Medium: 25-30\%, High: 45-50\%).}
\begin{tabular}{ccccccccccc}
\hline 
\multirow{2}{*}{Event 1} & \multirow{2}{*}{Censoring} & \multirow{2}{*}{True} & \multirow{2}{*}{n} & \multicolumn{3}{c}{Bias\% $\widehat{AUC}_{A}(\tau)$} &  & \multicolumn{3}{c}{MSE$\times10^{-3}$ $\widehat{AUC}_{A}(\tau)$}\tabularnewline
\cline{5-7} \cline{9-11} 
 &  &  &  & proposed & IPCW.KM. & IPCW.Cox &  & proposed & IPCW.KM & IPCW.Cox\tabularnewline
\hline 
\multirow{4}{*}{70\%} & \multirow{2}{*}{Medium} & \multirow{2}{*}{0.702} & 300 & 0.267  & 2.709  & 0.668  &  & 1.277  & 1.576  & 1.335 \tabularnewline
 &  &  & 600 & -0.069  & 2.426  & 0.412  &  & 0.562  & 0.829  & 0.596 \tabularnewline
\cline{2-11} 
 & \multirow{2}{*}{High} & \multirow{2}{*}{0.693} & 300 & 0.251  & 4.718  & 1.216  &  & 1.542  & 2.463  & 1.945 \tabularnewline
 &  &  & 600 & 0.336  & 4.888  & 1.401  &  & 0.690  & 1.839  & 1.029 \tabularnewline
\hline 
\multirow{4}{*}{50\%} & \multirow{2}{*}{Medium} & \multirow{2}{*}{0.691} & 300 & 0.180  & 2.689  & 1.035  &  & 1.578  & 1.861  & 1.811 \tabularnewline
 &  &  & 600 & 0.094  & 2.474  & 0.899  &  & 0.703  & 0.985  & 0.843 \tabularnewline
\cline{2-11} 
 & \multirow{2}{*}{High} & \multirow{2}{*}{0.693} & 300 & -0.347  & 4.802  & 1.190  &  & 1.782  & 2.708  & 2.734 \tabularnewline
 &  &  & 600 & 0.008  & 4.890  & 1.313  &  & 0.857  & 1.970  & 2.133 \tabularnewline
\hline 
\multirow{4}{*}{30\%} & \multirow{2}{*}{Medium} & \multirow{2}{*}{0.685} & 300 & 0.613  & 2.621  & 1.148  &  & 1.669  & 1.963  & 1.800 \tabularnewline
 &  &  & 600 & 0.453  & 2.471  & 1.125  &  & 0.843  & 1.125  & 0.915 \tabularnewline
\cline{2-11} 
 & \multirow{2}{*}{High} & \multirow{2}{*}{0.687} & 300 & -0.444  & 3.525  & 0.742  &  & 2.029  & 2.590  & 2.558 \tabularnewline
 &  &  & 600 & 0.123  & 4.125  & 1.282  &  & 0.991  & 1.772  & 1.355 \tabularnewline
\hline 
 &  &  &  &  &  &  &  &  &  & \tabularnewline
 &  &  &  & \multicolumn{3}{c}{Bias\% $\widehat{AUC}_{B}(\tau)$} &  & \multicolumn{3}{c}{MSE$\times10^{-3}$ $\widehat{AUC}_{B}(\tau)$}\tabularnewline
\cline{5-7} \cline{9-11} 
 &  &  &  & proposed & IPCW.KM. & IPCW.Cox &  & proposed & IPCW.KM & IPCW.Cox\tabularnewline
\hline 
\multirow{4}{*}{70\%} & \multirow{2}{*}{Medium} & \multirow{2}{*}{0.662} & 300 & 0.181  & 3.931  & 0.603  &  & 1.610  & 2.231  & 1.733 \tabularnewline
 &  &  & 600 & -0.183  & 3.630  & 0.322  &  & 0.743  & 1.286  & 0.802 \tabularnewline
\cline{2-11} 
 & \multirow{2}{*}{High} & \multirow{2}{*}{0.663} & 300 & 0.193  & 6.244  & 1.211  &  & 1.812  & 3.330  & 2.358 \tabularnewline
 &  &  & 600 & 0.265  & 6.467  & 1.353  &  & 0.814  & 2.651  & 1.251 \tabularnewline
\hline 
\multirow{4}{*}{50\%} & \multirow{2}{*}{Medium} & \multirow{2}{*}{0.652} & 300 & 0.537  & 4.874  & 1.383  &  & 1.986  & 2.862  & 2.403 \tabularnewline
 &  &  & 600 & 0.353  & 4.542  & 1.154  &  & 0.898  & 1.738  & 1.156 \tabularnewline
\cline{2-11} 
 & \multirow{2}{*}{High} & \multirow{2}{*}{0.671} & 300 & -0.277  & 7.155  & 1.311  &  & 2.115  & 4.140  & 3.395 \tabularnewline
 &  &  & 600 & 0.214  & 7.323  & 1.401  &  & 1.016  & 3.346  & 2.802 \tabularnewline
\hline 
\multirow{4}{*}{30\%} & \multirow{2}{*}{Medium} & \multirow{2}{*}{0.671} & 300 & 0.997  & 4.794  & 1.467  &  & 2.205  & 3.153  & 2.512 \tabularnewline
 &  &  & 600 & 0.617  & 4.454  & 1.236  &  & 1.115  & 1.975  & 1.231 \tabularnewline
\cline{2-11} 
 & \multirow{2}{*}{High} & \multirow{2}{*}{0.676} & 300 & -0.227  & 5.541  & 0.913  &  & 2.264  & 3.572  & 3.054 \tabularnewline
 &  &  & 600 & 0.412  & 6.265  & 1.494  &  & 1.136  & 2.876  & 1.668 \tabularnewline
\hline 
\end{tabular}
\end{sidewaystable}

\begin{sidewaystable}
\centering{}
\caption{Simulation results of $\widehat{Brier}(\tau)$ for the proposed method and
IPCW methods under dependent censoring setting $(b)$. Setting: event
1 rate (70\%, 50\% and 30\%), censoring rate (Medium: 25-30\%, High:
45-50\%).}
\begin{tabular}{ccccccccccc}
\hline 
\multirow{2}{*}{Event 1} & \multirow{2}{*}{Censoring} & \multirow{2}{*}{True} & \multirow{2}{*}{n} & \multicolumn{3}{c}{Bias\% BS} &  & \multicolumn{3}{c}{MSE$\times10^{-3}$ BS}\tabularnewline
\cline{5-7} \cline{9-11} 
 &  &  &  & proposed & IPCW.KM & IPCW.Cox &  & proposed & IPCW.KM & IPCW.Cox\tabularnewline
\hline 
\multirow{4}{*}{70\%} & \multirow{2}{*}{Medium} & \multirow{2}{*}{0.199} & 300 & -0.434  & -4.019  & -0.935  &  & 0.133  & 0.207  & 0.146 \tabularnewline
 &  &  & 600 & -0.311  & -3.907  & -0.747  &  & 0.064  & 0.128  & 0.069 \tabularnewline
\cline{2-11} 
 & \multirow{2}{*}{High} & \multirow{2}{*}{0.182} & 300 & -0.387  & -6.949  & -1.165  &  & 0.166  & 0.339  & 0.191 \tabularnewline
 &  &  & 600 & 0.107  & -6.543  & -0.530  &  & 0.083  & 0.238  & 0.108 \tabularnewline
\hline 
\multirow{4}{*}{50\%} & \multirow{2}{*}{Medium} & \multirow{2}{*}{0.165} & 300 & -1.326  & -5.477  & -1.638  &  & 0.174  & 0.262  & 0.188 \tabularnewline
 &  &  & 600 & -0.979  & -5.048  & -1.266  &  & 0.092  & 0.166  & 0.099 \tabularnewline
\cline{2-11} 
 & \multirow{2}{*}{High} & \multirow{2}{*}{0.171} & 300 & -0.615  & -8.321  & -1.468  &  & 0.190  & 0.402  & 0.202 \tabularnewline
 &  &  & 600 & -0.923  & -8.610  & -1.463  &  & 0.092  & 0.315  & 0.121 \tabularnewline
\hline 
\multirow{4}{*}{30\%} & \multirow{2}{*}{Medium} & \multirow{2}{*}{0.150} & 300 & -0.566  & -3.731  & -0.777  &  & 0.189  & 0.228  & 0.196 \tabularnewline
 &  &  & 600 & -0.966  & -4.094  & -1.149  &  & 0.090  & 0.127  & 0.089 \tabularnewline
\cline{2-11} 
 & \multirow{2}{*}{High} & \multirow{2}{*}{0.132} & 300 & -0.840  & -6.320  & -0.954  &  & 0.213  & 0.286  & 0.223 \tabularnewline
 &  &  & 600 & -0.675  & -6.127  & -0.818  &  & 0.103  & 0.168  & 0.102 \tabularnewline
\hline 
\end{tabular}
\end{sidewaystable}

\begin{figure}[pb]
\centering{}\includegraphics[scale=0.1]{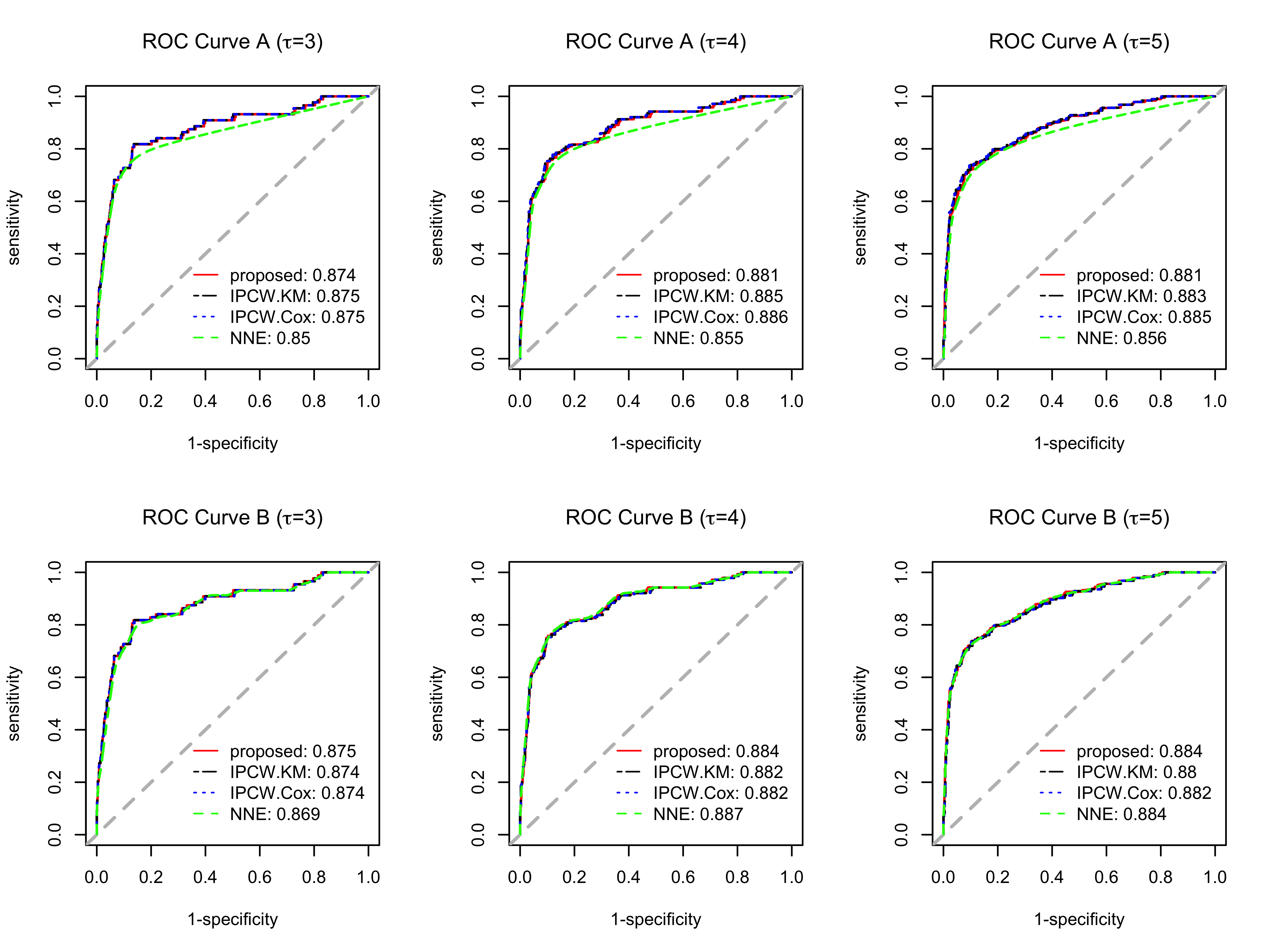}
\caption{$\widehat{ROC}(\tau)$ and $\widehat{AUC}(\tau)$ from the proposed method
(red), IPCW.KM (black), IPCW.Cox (blue) and NNE (green) evaluated at three prediction horizons: 3, 4 and 5 years from
baseline. ROC curve A corresponds to Definition A and ROC curve B
corresponds to Definition B. The $span=0.05$ was used. }
\end{figure}

\begin{figure}[pb]
\centering{}\includegraphics[scale=0.1]{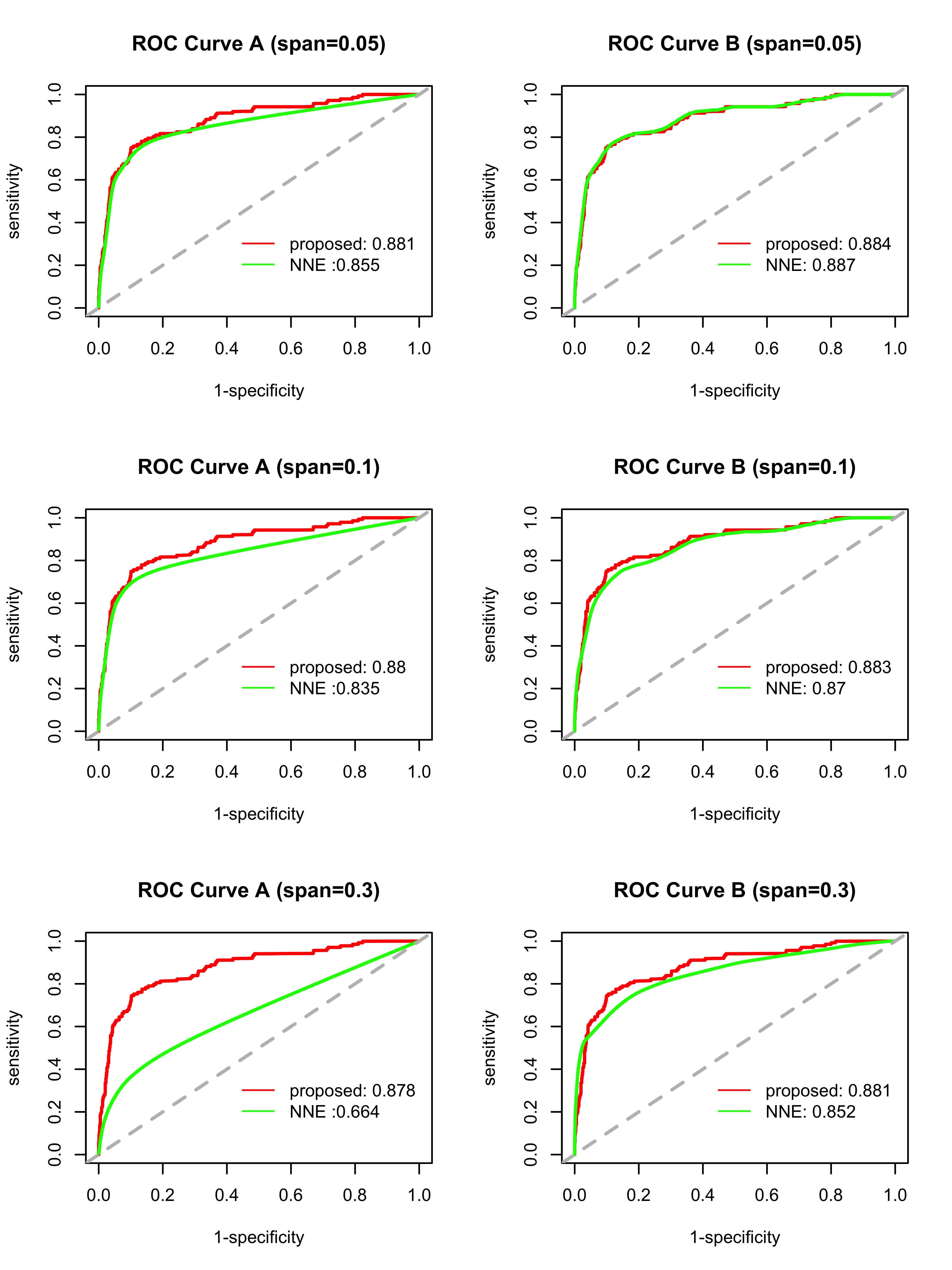}
\caption{$\widehat{ROC}(\tau)$ and $\widehat{AUC}(\tau)$ from the proposed method
(red) and NNE (green) using different $span$ evaluated at
year 4. ROC curve A corresponds to Definition A and ROC curve B corresponds
to Definition B. }
\end{figure}

\begin{figure}[pb]
\centering{}\includegraphics[scale=0.1]{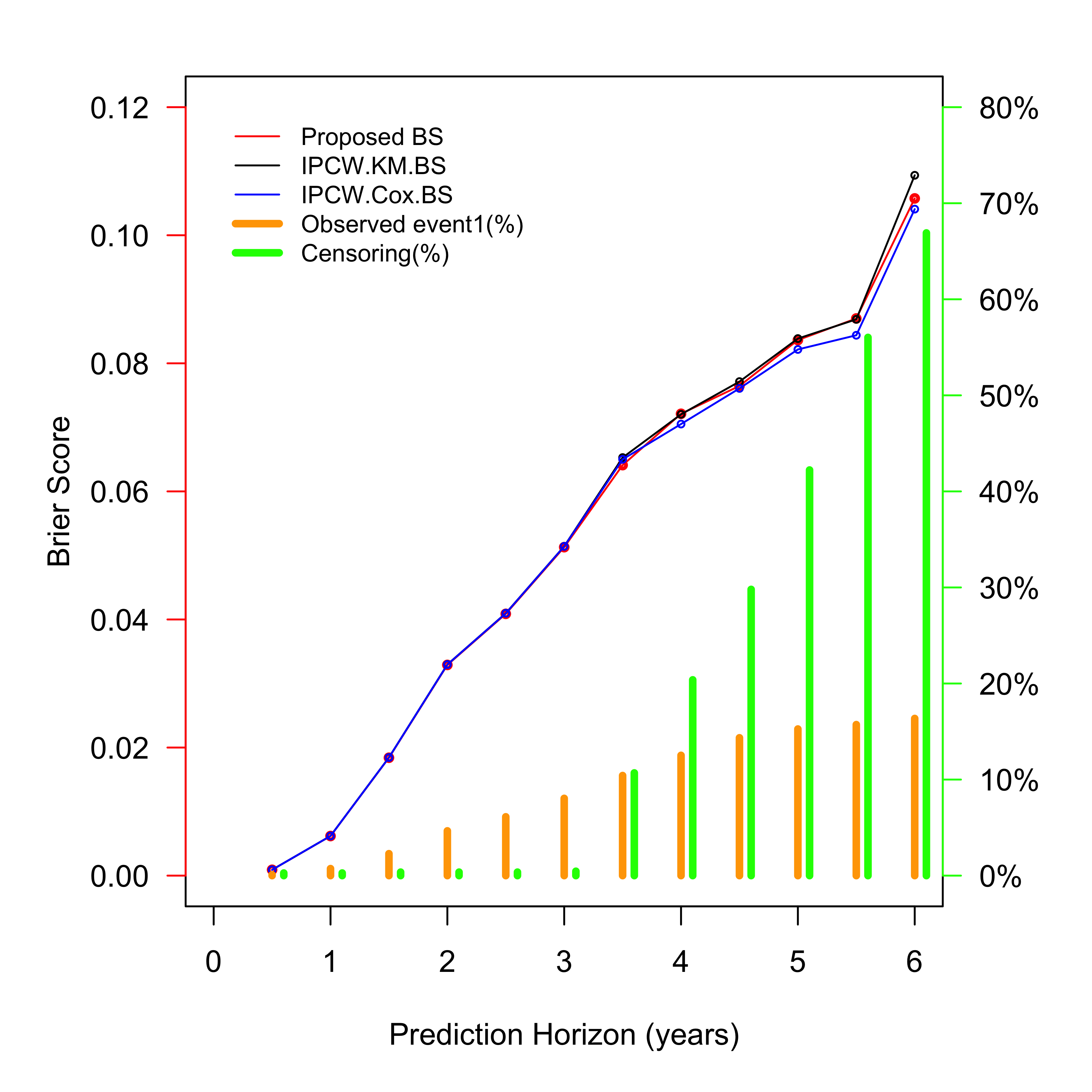}
\caption{$\widehat{Brier}(\tau)$ using proposed method (red), IPCW.KM (black)
and IPCW.Cox (blue). The orange bar indicates percent of ESRD and
green bar indicates percent of censoring prior to the prediction horizons, plotted
against the vertical axis on the right.}
\end{figure}
\pagebreak
\end{document}